\documentclass[english,british,1p]{elsarticle}
\usepackage{charter}
\usepackage[T1]{fontenc}
\usepackage[latin9]{inputenc}
\usepackage{amsmath}
\usepackage{graphicx}
\usepackage{babel}
\begin{document}
\title{Prediction of the vibro-acoustic response of a structure-liner-fluid system based on a patch transfer function approach and direct experimental subsystem characterisation}

\author[add1,add2]{Christopher G. Albert\corref{cor1}}
\ead{albert@alumni.tugraz.at}
\author[add1]{Giorgio Veronesi}
\author[add1]{Eugène Nijman}
\author[add1]{Jan Rejlek}

\cortext[cor1]{Corresp. author. Tel.: +43 3168738174; fax: +43 3168738677}
\address[add1]{Virtual Vehicle Research Center, Inffeldgasse 21A, A-8010 Graz, Austria}
\address[add2]{Institut für Theoretische Physik -- Computational Physics,\\ Technische Universität Graz, Petersgasse 16, A-8010 Graz, Austria}
\begin{abstract}
The vibro-acoustic response of a structure-liner-fluid system is predicted
by application of a patch transfer function (PTF) coupling scheme. In contrast
to existing numerical approaches, PTF matrices of structure
and liner are determined by a direct experimental approach, avoiding
the requirement of material parameters. Emphasis is placed on poroelastic
lining materials. The method accounts for surface input and next-neighbour transfer
terms and for cross and cross-transfer terms through the specimen.
Shear stresses and transfer terms to further patches on the liner are neglected.
A single test-rig characterisation procedure for layered poroelastic
media is proposed. The specimen is considered as a single component
-- no separation of layers is performed. For this reason the characterisation procedure
can serve as a complement to existing methods if separation of layers is not possible 
and as a tool for validation of more detailed material models.
Problem specific boundary conditions for skeleton and fluid, which may cause non-reciprocal
cross terms, are dealt with by the procedure. Methods of measurement
for the assessment of PTF matrices are presented and their accuracy
and limitations are discussed. An air gap correction method for surface
impedance measurements is presented.
\end{abstract}
\begin{keyword}
 Vibro-acoustic coupling \sep Impedance matrix \sep Patch transfer functions \sep Poroelastic material \sep \emph{PU}-probes
\end{keyword}
\maketitle

\section{Introduction}

Poroelastic lining materials are widely applied as dissipative treatments
in vibro-acoustic systems. When a structure is radiating into a cavity,
the insertion of poroelastic damping layers has three main effects:
Structural loading and damping, decoupling the cavity from the structure
(mass-spring systems) and adding absorption to the cavity.

In traditional numerical simulations \citep{Kropp_JCA_2003}, the
influence on the structure is usually described by additional mass
and damping. The damping imposed on the cavity by the poroelastic
layer is captured by an impedance boundary condition in the simplest
case. Required parameters may be estimated by material models, ranging
from simple equivalent mechanical systems to phenomenological impedance
models (e.g. \citep{Delany_AA_1970}). Mechanical parameters of lining
materials may be obtained in dynamic stiffness tests and impedances
are measured in a standing wave tube or in situ on the material surface
\citep{Lanoye_JASA_2006}. Depending on the type of lining material
and the desired accuracy, more detailed models are required. A common
approach is based on the Biot model for poroelasticity in a full FEM
simulation \citep{Atalla_JASA_2001} or in a reduced transfer matrix
scheme \citep{Allard_BOOK_2009}. Material parameters, such as porosity
or flow resistivity can be obtained experimentally on material samples.
Due to the variety of material properties (e.g. viscoelastic skeleton,
anisotropy, etc.), modelling and characterisation of poroelastic material
are an area of active research \citep{Jaouen_AA_2008}. 

The patch transfer function (PTF) coupling scheme \citep{Ouisse_JVA_2005,Aucejo_CS_2010,Pavic_CSV_2010}
has been introduced as a method to reduce the calculation time in
coupled fluid/fluid and fluid/structure simulations. While having
been developed for numerical applications, the relatively small number
of discrete surface elements (patches) makes this approach also applicable
to experimental characterisation of physical systems \citep{Rejlek_SAE_2013,Veronesi_SAE_2014}.
For structure-borne sound similar approaches have been introduced
to couple subsystems by mobility matrices (e.g. \citep{Petersson_AA_2000}). 

In this article the PTF methodology is applied to a physical structure/liner/fluid
system with experimentally characterised structure and liner. The
principles of the coupling method and the experimental realisation
for the subsystem characterisation are presented. Particular emphasis
is given to the liner characterisation method, which is non-de\-structive
and can be performed on flat and isotropic samples. Non-local effects
due to wave propagation in the lateral direction of the liner are
accounted for by transfer terms. Propagation across the thickness
of the material is described by cross terms. Mechanical separation
of the layers, with the risk of modifying their characteristics, is
not necessary. 

The described direct characterisation methods are intended to capture
the response of a subsystem as-is. For example if no reliable numerical
models are available for a complex structure such as a car body, experimentally
acquired patch mobilities may be used instead. A direct experimental
liner characterisation may not only be be of interest when no material
models are available, but also for materials where a separation of
layers is not possible or when parameters vary continuously across
the sample thickness. However, results from a direct experimental
characterisation do not allow for later adjustments in material parameters
or geometry. For these reasons, the approach is considered as \emph{complementary
to} rather than as \emph{a replacement for} micro-models based on
material parameters. Since patch transfer matrices can also be computed
from the latter, the proposed methods might be useful as an intermediate-step
experimental verification of modelling results. 

Characterisation results of isolated subsystems and the coupled system
are presented in section \ref{sec:Characterisation-results}. An overview
of the limitations of PTF coupling and characterisation methods can
be found in section \ref{sec:Limitations}. The range of validity
is roughly set by the frequency where the wavelength in any of the
coupled systems reaches the spatial aliasing limit. High dynamic range
of sensors and highly accurate calibration and characterisation measurements
are required to avoid random and systematic errors masking the results.

\section{Theory}

\subsection{Patch transfer functions\label{sub:Patch-transfer-functions}}

\begin{figure}
\begin{centering}
\includegraphics{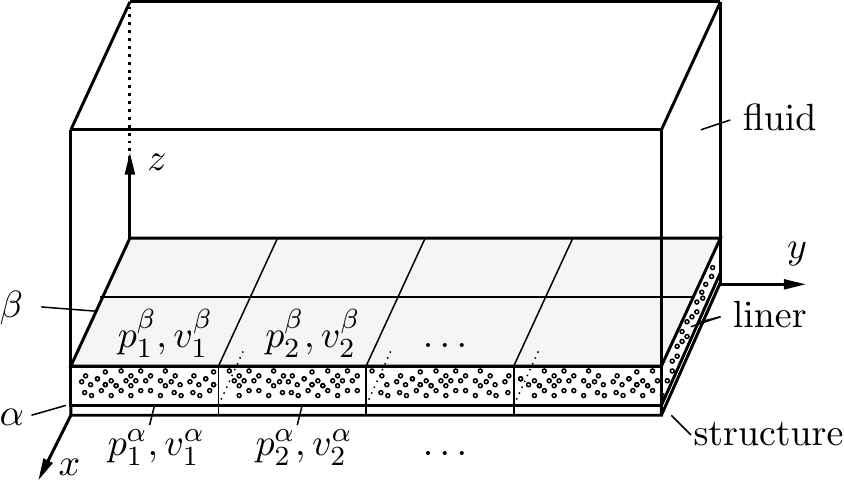}
\par\end{centering}

\protect\caption{\label{fig:Coupled-system-discretised}Coupled system discretised
into patches. The $\alpha$ and $\beta$ superscripts indicate the
structure/liner interface and the liner/fluid interface respectively.}
\end{figure}

This section introduces PTF coupling equations for a built-up system
consisting of a structure, liner and fluid domain. The system is discretised
by a patch grid as shown in Figure 1. The arithmetic mean of complex
field amplitudes is taken over each patch surface. Thereby, integral
equations for infinitesimal elements and Green's functions are approximated
by matrix equations for discrete patches. Detailed derivations of
the patch transfer function method can be found in \citep{Ouisse_JVA_2005,Pavic_CSV_2010,Bobrovnitskii_AP_2001}.
The present approach considers only out-of-plane velocities for the
coupling procedure without explicitly including shear stresses when
coupling to the liner. This is justified at least in the case of a
thin plate-like structure, where in-plane and out-of-plane velocity
are directly related and the directivity of the radiated wave is similar
with and without liner \citep{Allard_BOOK_2009}. If there is no direct
contact between structure and liner (a thin layer of air in-between)
it is also possible to neglect shear stresses \citep{Bolton_JSV_1996}.

In the following a slightly alternative matrix formulation is used
to describe coupling between sub-systems. Structure and fluid are,
as usual, characterised by respectively a mobility matrix $\mathbf{Y}$
and an impedance matrix $\mathbf{Z}$. Since patch-averaged pressures
$p_{i}$ and velocities $v_{i}$ are the governing variables, an impedance
or mobility term can be either interpreted as a patch-averaged acoustic
surface impedance/mobility or a mechanical impedance $Z_{{\rm mech}}/S_{p}$
or mobility $S_{p}Y_{{\rm mech}}$ normalised by a factor of the patch
area $S_{p}$ . The liner will be characterised by a hybrid matrix
\textbf{$\mathbf{H}$} instead of a conventional impedance matrix.
A similar technique has been described by \citet{Atalla_JASA_2001}
to integrate poroelastic materials into finite element models of structural
and acoustic domains.

The interface between the liner and the structure (respectively the
cavity) is referred to as face $\alpha$ (respectively face $\beta$,
see Figure \ref{fig:Coupled-system-discretised}). Positive velocities
are directed upwards by convention, which leads to a change of signs
in the mobility relation of the structure. The velocity response of
the structure due to a pressure excitation on face $\alpha$ is given
by
\begin{align}
\mathbf{v}^{\alpha} & =-\mathbf{Y}\mathbf{p^{\alpha}}.\label{eq:Y}
\end{align}
The mobility $\mathbf{Y}$ is an intrinsic property of the structure,
so relation (\ref{eq:Y}) holds independently from the actual source
of the pressure -- whether it is due to an external excitation or
due to the coupled response of another subsystem. If an excitation
of $p_{n}^{\alpha}=1$ at position $n$ of an otherwise free ($p_{m\neq n}^{\alpha}=0$)
structure is applied, the resulting response velocities are the elements
of the $n$-th row of $\mathbf{Y}$. 

The fluid surface impedance on $\beta$ relates pressures and velocities
on the upper liner-fluid coupling surface,
\begin{align}
\mathbf{p}^{\beta} & =\mathbf{Z}\mathbf{v}^{\beta}.
\end{align}
This relation is again independent of the coupling and the $n$-th
row of $\mathbf{Z}$ is equal to the hypothetically blocked ($v_{n\neq m}^{\beta}=0$
) pressures due to an excitation of $v_{n}^{\beta}=1$ at position
$n$.

A relationship between pressure and velocity on the two surfaces of
the liner is given by the matrix equation
\begin{align}
\left[\begin{array}{c}
\mathbf{p}^{\alpha}\\
\mathbf{v}^{\beta}
\end{array}\right] & =\mathbf{H}\left[\begin{array}{c}
\mathbf{v}^{\alpha}\\
\mathbf{p}^{\beta}
\end{array}\right]=\left[\begin{array}{cc}
\mathbf{h}^{\alpha\alpha} & \mathbf{h}^{\alpha\beta}\\
\mathbf{h}^{\beta\alpha} & \mathbf{h}^{\beta\beta}
\end{array}\right]\left[\begin{array}{c}
\mathbf{v}^{\alpha}\\
\mathbf{p}^{\beta}
\end{array}\right].\label{eq:Hmatrix}
\end{align}
$\mathbf{v}^{\alpha}$ describes a velocity (kinematic) excitation
from the bottom side and $\mathbf{p}^{\beta}$ a pressure excitation
from the top side.

For the one-dimensional case (one patch on each side), bottom and
top quantities $\mathbf{p}^{i},\,\mathbf{v}^{i}$ and sub-matrices
$\mathbf{h}^{ij}$ are given by scalars $p^{i},\,v^{i},\,h^{ij}$
that may be interpreted in the following way:

\begin{align}
h^{\alpha\alpha} & =\left.\frac{p^{\alpha}}{v^{\alpha}}\right|_{p^{\beta}=0}
\end{align}
 is the impedance as seen from the bottom side while keeping the top
side free (the inverse of the bottom mobility). 

\begin{align}
h^{\alpha\beta} & =\left.\frac{p^{\alpha}}{p^{\beta}}\right|_{v^{\alpha}=0}
\end{align}
 is the transmission ratio from a pressure excitation on the top to
the blocked bottom.

\begin{align}
h^{\beta\alpha} & =\left.\frac{v^{\beta}}{v^{\alpha}}\right|_{p^{\beta}=0}
\end{align}
 is the transmission ratio from a velocity excitation on the bottom
to the free top.

\begin{align}
h^{\beta\beta} & =\left.\frac{v^{\beta}}{p^{\beta}}\right|_{v^{\alpha}=0}
\end{align}
 is the top surface mobility (the inverse of the surface impedance)
with a blocked bottom.

The $\mathbf{H}$-matrix elements can in principle be obtained by
straightforward measurement of the state variables, provided well
defined boundary conditions are imposed. As will become clear in section
\ref{sub:Liner-characterisation--} it is actually not necessary to
measure the matrix elements under idealised conditions (velocity blocked
at the bottom interface, pressure release at the top interface), but
rather reconstruct them by solving an inverse problem for conditions
similar to the ideal case.

Let $\tilde{\mathbf{v}}^{\alpha}$ and $\tilde{\mathbf{p}}^{\alpha}$
be the source terms, i.e. the responses of the isolated systems without
liner to\emph{ internal} excitations (not on $\alpha$ or $\beta$).
In other words, $\tilde{\mathbf{v}}^{\alpha}$ is the response of
the free ($\tilde{\mathbf{p}}^{\alpha}=\mathbf{0}$) structure due
to some internal structural excitation and $\tilde{\mathbf{p}}^{\beta}$
the response of the blocked ($\tilde{\mathbf{v}}^{\beta}=\mathbf{0}$)
fluid due to some internal acoustical excitation. Then, using the
superposition principle \citep{Bobrovnitskii_AP_2001}, the response
of the coupled system is given by the following set of equations:
\begin{align}
\mathbf{v}^{\alpha} & =\tilde{\mathbf{v}}^{\alpha}-\mathbf{Y}\mathbf{p}^{\alpha},\\
\mathbf{p}^{\alpha} & =\mathbf{h}^{\alpha\alpha}\mathbf{v}^{\alpha}+\mathbf{h}^{\alpha\beta}\mathbf{p}^{\beta},\\
\mathbf{v}^{\beta} & =\mathbf{h}^{\beta\alpha}\mathbf{v}^{\alpha}+\mathbf{h}^{\beta\beta}\mathbf{p}^{\beta},\\
\mathbf{p^{\beta}} & =\tilde{\mathbf{p}}^{\beta}+\mathbf{Z}\mathbf{v}^{\beta},
\end{align}

or in matrix form
\begin{align}
\left[\begin{array}{cccc}
\mathbf{I} & \mathbf{Y} & 0 & 0\\
-\mathbf{h}^{\alpha\alpha} & \mathbf{I} & 0 & -\mathbf{h}^{\alpha\beta}\\
-\mathbf{h}^{\beta\alpha} & 0 & \mathbf{I} & -\mathbf{h}^{\beta\beta}\\
0 & 0 & -\mathbf{Z} & \mathbf{I}
\end{array}\right]\left[\begin{array}{c}
\mathbf{v}^{\alpha}\\
\mathbf{p}^{\alpha}\\
\mathbf{v}^{\beta}\\
\mathbf{p}^{\beta}
\end{array}\right] & =\left[\begin{array}{c}
\tilde{\mathbf{v}}^{\alpha}\\
0\\
0\\
\tilde{\mathbf{p}}^{\beta}
\end{array}\right].\label{eq:PTF}
\end{align}

For the case without liner where surfaces $\alpha$ and $\beta$ coincide,
the corresponding reduced $\mathbf{H}$-matrix is given by 
\begin{align}
\mathbf{H} & =\left[\begin{array}{cc}
\mathbf{0} & \mathbf{I}\\
\mathbf{I} & \mathbf{0}
\end{array}\right]
\end{align}
and (\ref{eq:PTF}) is reduced to 
\begin{align}
\left[\begin{array}{cc}
\mathbf{I} & \mathbf{Y}\\
\mathbf{-Z} & \mathbf{I}
\end{array}\right]\left[\begin{array}{c}
\mathbf{v}\\
\mathbf{p}
\end{array}\right] & =\left[\begin{array}{c}
\tilde{\mathbf{v}}\\
\tilde{\mathbf{p}}
\end{array}\right],\label{eq:PTFbare}
\end{align}
where $\mathbf{v}^{\alpha}=\mathbf{v}^{\beta}=\mathbf{v}$ and $\mathbf{p}^{\alpha}=\mathbf{p}^{\beta}=\mathbf{p}$.

\section{Subsystem characterisation}

$\mathbf{Y},$ \textbf{$\mathbf{Z}$ }and $\mathbf{H}$ may be obtained
by either numerical or experimental means. In existing works, PTF
matrices were derived by analytical \citep{Ouisse_JVA_2005,Pavic_CSV_2010}
or numerical \citep{Aucejo_CS_2010} means. In the following sections
some techniques for the direct experimental characterisation of subsystems
without the requirement of a material numerical model are proposed.
In the reconstruction these subsystems are combined with each other
or with their numerical counterparts to predict the response of the
coupled system.

The principle of the proposed characterisation method consists in
the measurement of the isolated subsystem response for a number of
load cases equal to the number of patches, thereby forming a full
set of linear equations. The effort required for the experimental
characterisation of a subsystem scales with the square of the number
of surface patches, since the transfer function is measured between
each pair of patches. The required time to characterise a grid of
32 patches on a structure is in the order of 2-3 working days. The
characterisation of the next-neighbour $\mathbf{H}$-matrix of a liner
specimen is usually performed within half a working day. For fluid
cavities, no reliable data is available yet.

\subsection{Structure\label{sub:Structure}}

A structure with a relatively high impedance is assumed to be freely
vibrating in air ($Y^{-1}\gg Z_{0})$. Therefore, it is possible to
measure the terms $y_{ij}$ of the mobility matrix in a direct way
by exciting on position $j$ and measuring the velocity on position
$i$ . In our case the excitation was applied by means of an impact
hammer. An excitation similar to a uniform pressure excitation was
realised by averaging over a sufficient number of hammer blows on
an equidistant grid inside a patch. If the grid point distance is
small compared to the structural wavelength, the excitation corresponds
to a uniform pressure excitation of the whole patch. Convergence to
this condition can be proven theoretically by wavelength criteria
and experimentally by reaching the level of grid refinement where
the results stop changing significantly in the desired frequency range.
The response can be measured, as in our case, by a grid of accelerometers,
a scanning laser vibrometer or by an array of $PU$-probes. In practice,
convergence was reached by a $4\times4$ grid of excitation and receiver
positions per patch for the plate described in section \ref{sub:Plate-characterisation},
which has been confirmed by a Kirchhoff-Love model.

\subsection{Fluid}

A realisation of an experimental cavity characterisation is currently
under investigation but has not been performed yet. For a direct characterisation
of the fluid cavity a flat, square-shaped piston source should in
principle be progressively positioned in the different patch locations.
As an alternative to this piston-like radiator, a superposition of
excitations by a point source (\textquotedblleft tube\textquotedblright{}
source) might be possible, analogous to the point excitation of the
structure with a hammer. The blocked (the cavity must be equipped
with a high impedance wall) pressure response $z_{ij}$ would be measured
using an array of pressure microphones or $PU$-probes, where $PU$-probes
offer the advantage of directly measuring the particle velocity on
the exciting patch.

\subsection{Liner characterisation -- fundamentals\label{sub:Liner-characterisation--}}

Acoustical liners feature a porous solid phase (also referred to as
skeleton or matrix) with interconnected interstitial cavities filled
with a fluid phase (air). At the two coupling surfaces the area fraction
occupied by the solid phase is very small. At the structure/liner
interface, however, the considerable pressure concentrations in the
tiny contact areas between skeleton and adjacent structure cause fluid
and skeleton to have the same normal velocity. The characterisation
procedure must account for this feature and mechanical excitation
with a piston is thus necessary at this interface in order to obtain
representative data.

Moreover, at the structure/liner interface both solid as well as fluid
phase may significantly contribute to the average patch pressure,
and microphones can consequently not be used, but the average patch
pressure must be obtained indirectly through piston force measurements. 

At the liner/fluid interface, on the other hand, the fluid of the
cavity adjacent to the liner is unable to support shear stresses and
no pressure concentration will occur. Usually continuity of pressure,
normal stresses and normal displacements are assumed on the liner/fluid
interface~\citep{Debergue_JASA_1999}. In the coupling conditions,
displacements of fluid and solid phase of the poroelastic material 
are weighted depending on its porosity.
Consequently, normal velocities of fluid and solid phase do not 
necessarily coincide, as opposed to the structure/liner interface.
The characterisation procedure should therefore use an acoustic
excitation together with microphones and particle velocity sensors
at the liner/fluid interface, while avoiding a direct mechanical 
excitation there which would kinematically couple normal displacements 
of solid and fluid phase.

\begin{figure}
\begin{centering}
\includegraphics{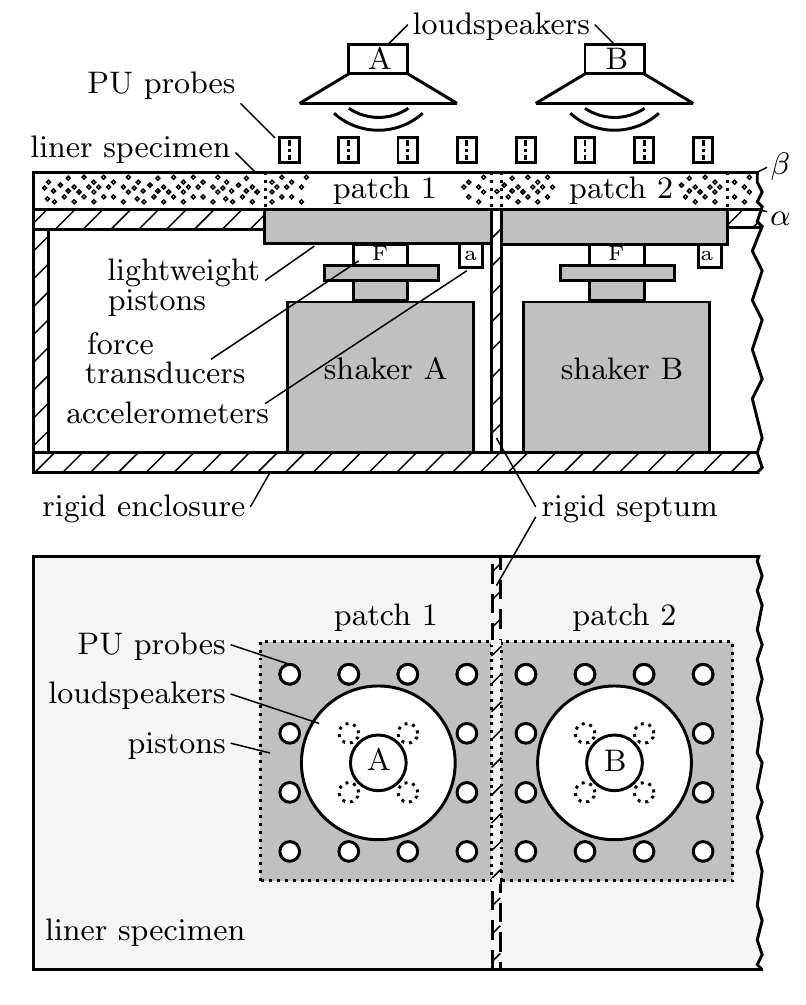}
\par\end{centering}

\protect\caption{Liner test rig for two patches in cross-sectional view (top) and from
above (bottom). Load cases are imposed by loudspeakers A, B and shakers
A, B via square pistons. The response is measured by PU probes, accelerometers
and force transducers. Enclosure and septum avoid acoustic leakage
and the latter is thin compared to the patch width. \label{fig:Principle-of-the}}
\end{figure}

To experimentally re-construct the $\mathbf{H}$ matrix on the $2N$
patches (top and bottom), $2N$ linearly independent load cases are
needed while measuring pressure and velocity on all patches. These
load cases are imposed by structural shaker excitations on the lower
surface $\alpha$ and speaker excitations on the upper surface $\beta$.
The measurement of pressures and velocities on top and bottom side
is performed by a combination of force transducers, accelerometers
on the bottom side and \emph{PU}-probes on the top side. In principle
one shaker and one speaker is needed for each patch excitation. A
sketch of the test rig is displayed in Figure \ref{fig:Principle-of-the}.
Stable results for two patches were obtained using the following four
excitation conditions as a sweep or noise signals in the considered
frequency range:\medskip{}
\\
$I:$ loudspeakers \textquotedblleft off\textquotedblright{} and shaker
A \textquotedblleft on\textquotedblright ,\\
$II:$ loudspeakers \textquotedblleft off\textquotedblright{} and
shaker B ``on'',\\
$III:$ shakers \textquotedblleft off\textquotedblright{} and speakers
A and B \textquotedblleft on\textquotedblright{} in phase, \\
$IV:$ shakers \textquotedblleft off\textquotedblright{} and speakers
A and B \textquotedblleft on\textquotedblright{} in anti-phase. \medskip{}
\\
For the air gap correction (section \ref{sub:Air-gap-correction})
load cases $III$ and $IV$ are applied once more with the liner specimen
replaced by a rigid dummy.

\subsubsection{Bottom surface active patch}

The bottom surface of the liner is excited by a shaker on patch $i$.
To excite the whole patch, a patch-sized light\-weight piston is
mounted on top of the shaker. The velocity of the piston, $v_{i}^{\alpha}$,
is measured using an accelerometer. Since a direct pressure measurement
on the interface between piston and liner is difficult to achieve,
the reaction force $F_{i}^{\alpha,\,{\rm meas}}$ below the piston
is measured instead by force cells. This force includes the response
of both piston and liner. In the frequency range, where the piston
moves as a rigid body, the reaction pressure on the liner bottom surface
is given by 
\begin{align}
p_{i}^{\alpha} & =F_{i}^{\alpha,\,{\rm meas}}/S_{p}-Z_{i}^{act}v_{i}^{\alpha},\label{eq:p1a}
\end{align}
where $Z_{i}^{{\rm act}}$ is the piston's acoustic impedance and
$S_{p}$ is the area of one patch.

For a rigid piston, 
\begin{equation}
Z_{i}^{{\rm act}}=Z_{{\rm m},\,i}^{{\rm act}}+Z_{{\rm rad},\,i}^{{\rm act}}
\end{equation}
is governed by a mass-like mechanical impedance normalised to the
patch area, $Z_{{\rm m},\,i}^{{\rm act}}=i\omega m_{i}/S_{p}$ and
the acoustic radiation impedance $Z_{{\rm rad},\,i}^{{\rm act}}$.
It is determined by a calibration measurement consisting of driving
the plate without liner. Like in the later characterisation measurement,
the force $F_{i}^{\alpha,\,{\rm cal}}$ below the piston is measured
by the force transducers and the pressure $p_{i}^{\alpha,\,{\rm cal}}$
flush to the piston surface as well as the piston velocity $v_{i}^{\alpha,\,{\rm cal}}$
using a \emph{PU}-probe array. The \emph{PU} particle velocity sensor
sensitivities are adjusted to match the piston accelerometer reading.
The piston's acoustical impedance on patch $i$ is obtained as
\begin{align}
Z_{i}^{{\rm act}}= & \frac{F_{i}^{\alpha,\,{\rm cal}}/S_{p}-p_{i}^{\alpha,\,{\rm cal}}}{v_{i}^{\alpha,\,{\rm cal}}}.\label{eq:Zact}
\end{align}

\subsubsection{Bottom surface passive patches\label{sub:Bottom-surface-passive}}

For load cases requiring a non-driven bottom patch $j$, the pressure
on the plate surface is determined indirectly by the relation
\begin{align}
p_{j}^{\alpha} & =Z_{j}^{{\rm pass}}v_{j}^{\alpha}.
\end{align}
Here the acoustical impedance $Z_{j}^{{\rm pass}}$ of the passive
plate is again obtained through a calibration measurement. The speaker
on patch $j$ is switched on and the calibration pressure $p_{j}^{\alpha,\,{\rm cal}}$
is again determined from measurements with the \emph{PU} array flush
to the surface. The impedance of the passive system (combination of
plate and remaining set-up below) is then given directly as 
\begin{align}
Z_{j}^{{\rm {\rm pass}}} & =\frac{p_{j}^{\alpha,\,{\rm cal}}}{v_{j}^{\alpha,\,{\rm cal}}}\,.
\end{align}

\subsubsection{Top surface}

Acoustical excitation is provided by loudspeakers cen\-tred above
each patch as shown in Figure \ref{fig:Principle-of-the}. Pressures
and velocities are measured by averaging over an array of \emph{PU}-probes
located between the loudspeaker and the liner as close as possible
to the surface of the latter.

\subsubsection{Reconstruction of the H-matrix}

According to (\ref{eq:Hmatrix}), the $\mathbf{H}$-matrix of the
liner can be reconstructed from $2N$ load cases by
\begin{align}
\mathbf{H} & =\left[\begin{array}{c}
\mathbf{P}^{\alpha}\\
\mathbf{V}^{\beta}
\end{array}\right]\left[\begin{array}{c}
\mathbf{V}^{\alpha}\\
\mathbf{P}^{\beta}
\end{array}\right]^{-1},
\end{align}
where the matrices $\mathbf{P}^{\alpha},\,\mathbf{P}^{\beta},\,\mathbf{V}^{\alpha},\,\mathbf{V}^{\beta}$
contain the excitation and response pressures and velocities for each
load case (labelled by Roman numberals $I,II,\dots$) in one column,
i.e.
\begin{align}
\left[\begin{array}{c}
\mathbf{V}^{\alpha}\\
\mathbf{P}^{\beta}
\end{array}\right] & =\left[\left[\begin{array}{c}
\mathbf{v}^{\alpha}\\
\mathbf{p}^{\beta}
\end{array}\right]_{I}\left[\begin{array}{c}
\mathbf{v}^{\alpha}\\
\mathbf{p}^{\beta}
\end{array}\right]_{II}\dots\left[\begin{array}{c}
\mathbf{v}^{\alpha}\\
\mathbf{p}^{\beta}
\end{array}\right]_{2N}\right].
\end{align}
In principle any linearly independent set of excitation conditions
is suited to reconstruct the matrix H. In practice, however, bad conditioning
of the matrix $\left[\begin{array}{c}
\mathbf{V}^{\alpha}\\
\mathbf{P}^{\beta}
\end{array}\right]$ may lead to substantial error amplification in the inversion process,
especially when experimental data containing inevitable inaccuracies
have to be processed. Ideal excitation conditions consist of an orthonormal
set of vectors $\left[\begin{array}{c}
\mathbf{v}^{\alpha}\\
\mathbf{p}^{\beta}
\end{array}\right]_{i}$. In the following section a simplification will be introduced to
reduce the number of load cases using symmetry properties.

\subsection{Liner characterisation -- further remarks\label{sub:Liner-characterisation---}}

\subsubsection{Decay and Symmetries}

Wave propagation inside the liner in the lateral (in-plane) direction
is characterised by spatial decay. In the particular case of a moulded
foam with a limp heavy layer considered in this article only the driven
patch itself and the patch adjacent to the driven patch need to be
considered, whereas propagation to farther patches can be neglected.
Such an assumption can be checked in the following ways.

When approximate values of material data such as stiffness or damping
are known, transfer matrix terms to further patches can be estimated
from a corresponding material model. In the particular case considered
in this article, a numerical model of an equivalent elastic material
has been constructed and updated using the characterisation results.
Another option consists in measuring further transfer terms by adjustable
patch positions in the test rig. Once the additional terms do not
change the reconstruction significantly, the remaining transfer functions
to further patches are not characterised and set to zero in the calculation.

$\mathbf{H}$ may consequently be approximated by a banded matrix
and the test rig be limited to two patches in the present case. For
two adjacent patches \textbf{$1$ }and $2$, pressures and velocities
are labelled $p_{1}^{i}$, $p_{2}^{i}$, \textbf{$v_{1}^{i}$ }and
\textbf{$v_{2}^{i}$}. The elements of the $2\times2$ sub-matrices
of a banded $\mathbf{H}$ matrix in equation (\ref{eq:Hmatrix}) are
\begin{align}
h_{(12)}^{ij} & =\left[\begin{array}{cc}
h_{11}^{ij} & h_{12}^{ij}\\
h_{21}^{ij} & h_{22}^{ij}
\end{array}\right],
\end{align}
with surface $i,j\in\{\alpha,\beta\}$. If the material is both homogeneous
and isotropic in the lateral direction, the corresponding sub-matrices
$h_{(ab)}^{ij}$ for any two adjacent patches $a$ and $b$ are identical
and symmetric. The single elements of $h_{(ab)}^{ij}=h_{(12)}^{ij}=:\,h^{ij}$
are called input (in) and transfer (tr) terms, where
\begin{align}
h^{ij} & =\left[\begin{array}{cc}
h_{{\rm in}}^{ij} & h_{{\rm tr}}^{ij}\\
h_{{\rm tr}}^{ij} & h_{{\rm in}}^{ij}
\end{array}\right].\label{eq:h_in_tr}
\end{align}
The input element $h_{{\rm in}}^{ij}=h_{11}^{ij}=h_{22}^{ij}$ describes
the input response and the transfer element $h_{{\rm tr}}^{{\rm ij}}=h_{12}^{ij}=h_{21}^{ij}$
the response of the next neighbour. 

This symmetry leads to a reduction of the required load cases in the
characterisation measurements. Let $I$ denote a load case where an
excitation of $v_{1}^{\alpha}$ is applied on patch $1$ and $v_{2}^{\alpha}$
on patch $2$. In load case $II$ identical excitations are applied
but the patches are swapped, so $v_{2}^{\alpha}$ is applied on patch
$1$ and $v_{1}^{\alpha}$ on patch $2$. Making use of the symmetry
properties of $h^{ij}$, one can identify 
\begin{align}
\left[\begin{array}{c}
p_{1}^{\alpha}\\
p_{2}^{\alpha}
\end{array}\right]_{II} & =\left[\begin{array}{cc}
h_{{\rm in}}^{\alpha\alpha} & h_{{\rm tr}}^{\alpha\alpha}\\
h_{{\rm tr}}^{\alpha\alpha} & h_{{\rm in}}^{\alpha\alpha}
\end{array}\right]\left[\begin{array}{c}
v_{1}^{\alpha}\\
v_{2}^{\alpha}
\end{array}\right]_{II}+\dots\nonumber \\
 & =\left[\begin{array}{cc}
h_{{\rm in}}^{\alpha\alpha} & h_{{\rm tr}}^{\alpha\alpha}\\
h_{{\rm tr}}^{\alpha\alpha} & h_{{\rm in}}^{\alpha\alpha}
\end{array}\right]\left[\begin{array}{c}
v_{2}^{\alpha}\\
v_{1}^{\alpha}
\end{array}\right]_{I}+\dots=\left[\begin{array}{c}
p_{2}^{\alpha}\\
p_{1}^{\alpha}
\end{array}\right]_{I}.
\end{align}

\begin{figure}
\begin{centering}
\includegraphics{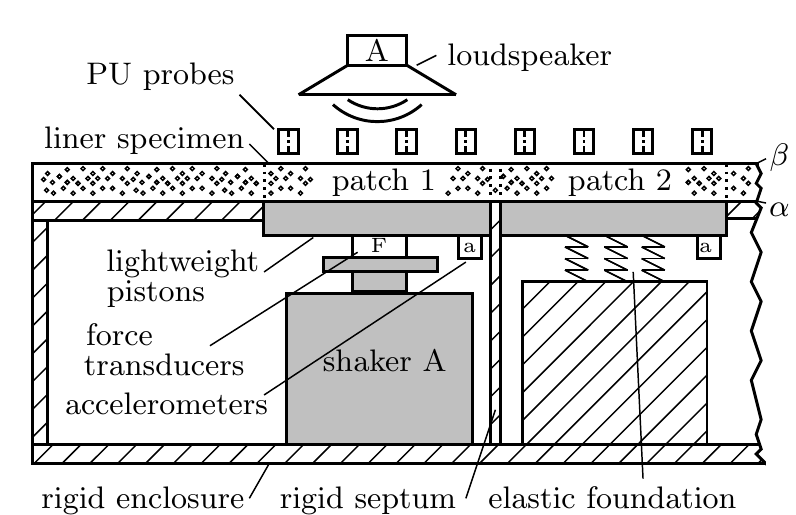}
\par\end{centering}

\protect\caption{Reduced liner test rig with one active patch excited by a shaker and
one passive patch with an elastic foundation.\label{fig:Reduced-liner-test}}
\end{figure}

Similar identities hold for the remaining possible excitation and
response quantities, which leads to a reduction of the required load
cases by half. The remaining (virtual) load cases are introduced by
swapping patch indices. This leads to the possibility of a reduced
version of the test rig with excitation only on patch 1 (Figure \ref{fig:Reduced-liner-test}).
In this set-up the piston below patch 2 is bedded on an elastic foundation
(e.g. springs or elastic foam). The pressure response below patch
2 is measured indirectly by the piston velocity as described in subsection
\ref{sub:Bottom-surface-passive}. The bedding should be chosen such
that translational and rotational mass-spring resonances are well
below the measurement range.

\subsubsection{Uncertainties in the indirect measurements}

It has to be noted that $Z_{i}^{{\rm act}}$ in (\ref{eq:p1a}) is
not necessarily the actual impedance of the free plate, but rather
a calibration factor that compensates for the combination of the plate
impedance and the calibration between force cell and microphones.
From equation (\ref{eq:p1a}), the uncertainty $\Delta p_{i}^{\alpha}$
of the liner reaction pressure on the active patch $i$ due to the
uncertainty $\Delta F_{i}^{\alpha,\,{\rm meas}}$ of the measured
force $F_{i}^{\alpha,\,{\rm meas}}$ is 
\begin{align}
\Delta p_{i}^{\alpha} & =\Delta F_{i}^{\alpha,\,{\rm meas}}/S_{p}\nonumber \\
 & =\frac{\Delta F_{i}^{\alpha,\,{\rm meas}}}{F_{i}^{\alpha,\,{\rm meas}}}\frac{F_{i}^{\alpha,\,{\rm meas}}}{S_{p}}\nonumber \\
 & =\frac{\Delta F_{i}^{\alpha,\,{\rm meas}}}{F_{i}^{\alpha,\,{\rm meas}}}\left(p_{i}^{\alpha}+Z_{i}^{act}v_{i}^{\alpha}\right).\label{eq:pia}
\end{align}
For nearly blocked bottom and nearly open top, one can assume 
\begin{align}
p_{i}^{\alpha} & \approx h_{{\rm in}}^{\alpha\alpha}v_{i}^{\alpha}.\label{eq:piapprox}
\end{align}
Using this relation in (\ref{eq:pia}) leads to an estimate of the
absolute error of
\begin{align*}
\Delta p_{i}^{\alpha} & \approx\frac{\Delta F_{i}^{\alpha,\,{\rm meas}}}{F_{i}^{\alpha,\,{\rm meas}}}\left(1+\frac{Z_{i}^{act}}{h_{{\rm in}}^{\alpha\alpha}}\right)p_{i}^{\alpha}.
\end{align*}

The relative error in the reconstructed liner pressure is thus amplified
as compared to the relative error in the direct force measurements.
The amplification factor is governed by the ratio of piston and liner
input impedance and thus, to a large extent, by the piston mass. The
piston must consequently not only behave like a rigid body in the
frequency range of interest but also be as lightweight as possible
in order to minimise the measurement uncertainty.

For the measurement on a passive plate $j$, a lightweight plate ensures
a higher velocity signal
\begin{align}
v_{j}^{\alpha} & \approx\frac{h_{{\rm tr}}^{\alpha\alpha}}{Z_{j}^{{\rm {\rm pass}}}}
\end{align}
above the noise level.

\subsection{Air gap correction\label{sub:Air-gap-correction}}

\begin{figure}
\begin{centering}
\includegraphics{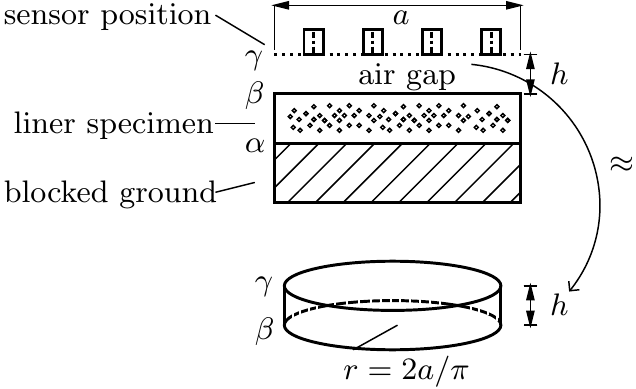}
\par\end{centering}

\protect\caption{Air gap of height $h$ and equivalent air cylinder.\label{fig:Airgap-of-heigth}}
\end{figure}

In principle the \emph{PU}-probes should be located in correspondence
with the top surface of the liner ($\beta$-surface), but in practice
a slight air gap between the liner surface and the plane of the transducers
is unavoidable. This air gap results in two error sources. On one
hand the air layer may exhibit radiation through the side walls of
the gap, on the other hand, the stiffness of the layer itself is finite.

The correction that will be applied for the rectangular patches is
based on a direct experimental characterisation of the air gap that
will be described in the following subsection. The qualitative behaviour
of the air gap can be illustrated by considering a thin cylindrical
air gap (Figure \ref{fig:Airgap-of-heigth}) of the same circumference
as a patch.

To estimate the acoustic input impedance $Z_{{\rm gap}}$ of the air
gap, the sample in the figure is replaced by a rigid surface. In the
frequency range where the wavelength is small in comparison to the
patch dimension, $Z_{{\rm gap}}$ is composed of the stiffness of
the air volume $V=\pi r^{2}h$ with bulk modulus $K_{{\rm air}}$
coupled in series to a radiation impedance $Z_{{\rm rad}}$ at the
outer rim. A uniform velocity excitation $v^{\gamma}$ at the upper
boundary $\gamma$ of the air gap leads to a compression of
\begin{align}
\Delta V & =\pi r^{2}\Delta h=\pi r^{2}\frac{v^{\gamma}}{i\omega},
\end{align}
so 
\begin{align}
Z_{{\rm gap}} & =\frac{p^{\gamma}}{v^{\gamma}}=\frac{r}{2h}\left(\frac{i\omega r}{2K_{{\rm air}}}+Z_{{\rm rad}}^{-1}\right)^{-1}.\label{eq:airgap}
\end{align}
The radiative term $Z_{{\rm rad}}$ will generally lead to a mass-like
drop in the low frequency limit \citet{Robey_JASA_1955}. The frequency
dependence of the stiffness term $i\omega r/(2K_{{\rm air}})$ leads
to another drop in $Z_{{\rm gap}}$ at higher frequencies. If the
liner surface input impedance is larger than $Z_{{\rm gap}}$, the
uncorrected characterisation result for this term is governed by the
air gap instead of the specimen, which will be quantified in the following
subsection. The power radiated through the rim of the gap towards
the neighbouring patches leads to an overestimation of transfer terms.
Experimental results confirming this behaviour will be presented in
section \ref{sub:Lining-material-characterisation}. The affected
frequency range depends on the liner type and the air-gap height $h$.
A lower height naturally leads to a higher air-gap impedance and to
improved characterisation results.

\subsubsection{Experimental air gap correction}

For an experimental correction procedure, the air gap is discretised
into the patch grid defined on the specimen surface. Since the thickness
of the air gap is very small as compared to the wavelength in the
frequency range of interest, the pressure may be considered uniformly
distributed within the air gap of each patch. The pressure vector
$\mathbf{p}^{\beta}$ measured at the top surface of the liner may
now be approximated by 
\begin{align}
\mathbf{p}^{\beta} & =\mathbf{Z}_{{\rm gap}}\left(\mathbf{v}^{\gamma}-\mathbf{v}^{\beta}\right)
\end{align}
where $\mathbf{v}^{\gamma}$ is the velocity vector containing the
patch velocities measured with the particle velocity transducers at
the upper surface of the air gap. $\mathbf{Z}_{{\rm gap}}$ represents
the air gap impedance matrix. The desired patch velocity vector $\mathbf{v}^{\beta}$
at the liner surface is obtained as
\begin{align}
\mathbf{v}^{\beta} & =\mathbf{v}^{\gamma}-\mathbf{Z}_{{\rm gap}}^{-1}\mathbf{p}^{\beta}.\label{eq:corr}
\end{align}
The air gap impedance matrix $\mathbf{Z}_{{\rm gap}}$ may be assessed
experimentally beforehand by replacing the liner specimen with a rigid
dummy ($\mathbf{v}^{\beta}=0$). By acquiring a sufficient number
of independent load cases the air gap impedance $\mathbf{Z}_{{\rm gap}}$
can again be obtained through a matrix inversion. Thus the particle
velocity at the liner specimen surface is reconstructed from the measurement
with \emph{PU} sensors at a finite distance from the specimen.

\section{Characterisation results\label{sec:Characterisation-results}}

In this section results of the experimental characterisation of a
metal plate and a liner are presented. Surfaces are discretised into
square patches of $20\times20\,{\rm cm}$. Displayed levels for characterisation
results are given in terms of impedance ($L_{Z}=20\log_{10}\left|\frac{p}{v}\right|$)
or mobility magnitudes ($L_{Y}=20\log_{10}\left|\frac{v}{p}\right|$)
with $p$ in ${\rm Pa}$ and $v$ in $m/s$, if not stated otherwise.

\subsection{Plate characterisation\label{sub:Plate-characterisation}}

\begin{figure}
\begin{centering}
\includegraphics{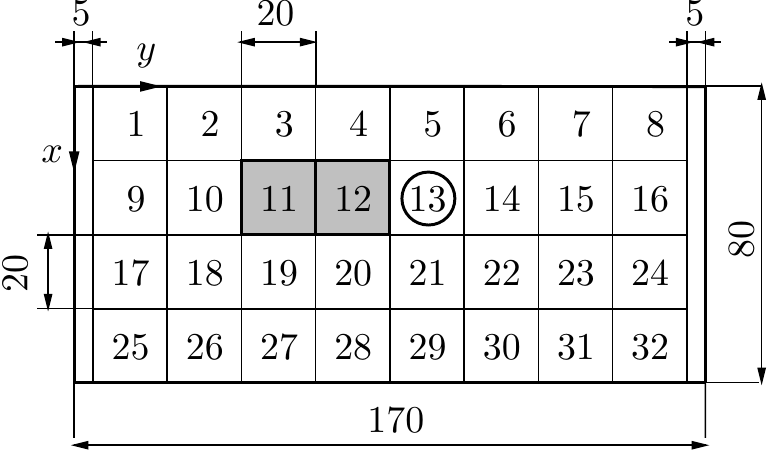}
\par\end{centering}

\protect\caption{Patch grid on a clamped plate with neglected strips to the left and
to the right (units: cm). The neighbouring patches $11$ and $12$
for which results are shown are marked in grey. Patch $13$, where
a point force is applied, is marked by a circle. \label{fig:Yplate-2}}
\end{figure}

\begin{figure}
\begin{centering}
\includegraphics[width=0.8\textwidth]{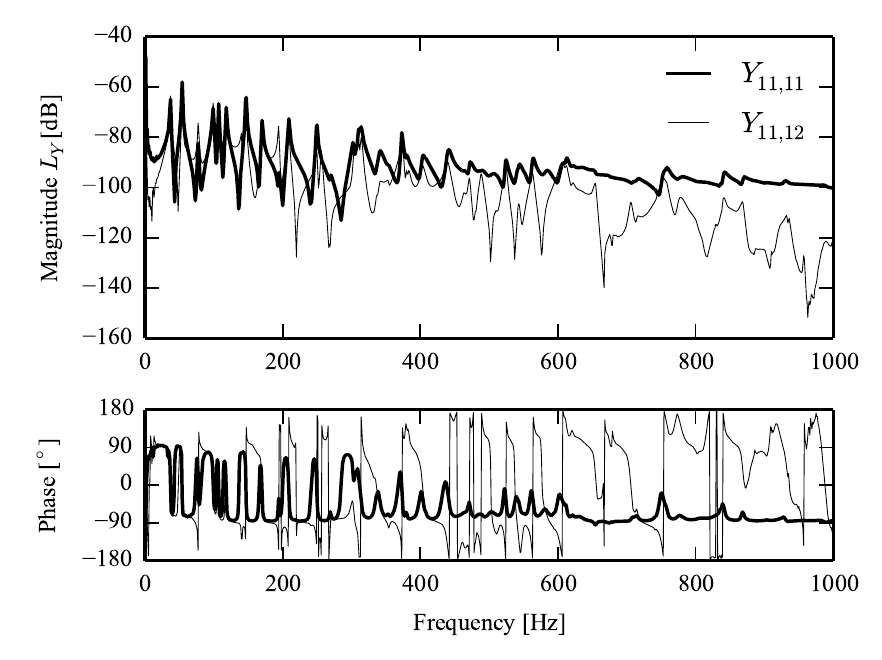}
\par\end{centering}

\protect\caption{Example of components of the plate mobility matrix $\mathbf{Y}$ --
an input term $Y_{11,11}$ and a transfer term $Y_{11,12}$ to the
neighbouring patch.\label{fig:Yplate}}
\end{figure}

A clamped aluminium plate of dimensions $1.7\times0.8\,{\rm m}$ and
a thickness of $5\,{\rm mm}$ (32 patches, Figure \ref{fig:Yplate-2})
has been characterised according to \ref{sub:Structure}. A strip
of $5\,{\rm cm}$ is neglected on two of the clamped edges, where
the velocity amplitude in the considered frequency range is very low.
Typical results for an input mobility $Y_{11,11}$ (excitation and
response on patch 11) and a transfer mobility $Y_{11,12}$ (excitation
on neighbouring patch 12) are shown in Figure \ref{fig:Yplate}.

For most of the lowest modes below 100 Hz, both amplitude and phase
of $Y_{11,11}$ and $Y_{11,12}$ are similar. This is expected since
the structural wave\-length is much longer than the patch size in
this frequency range. The average amplitude stays constant over frequency
which is the typical behaviour for the driving-point mobility of thin
plates \citep{Cremer_BOOK_2005}. In the region between 100 and 400
Hz the mobilities switch their phase relationship depending on the
respective mode and a phase run-off due to propagating waves is visible
in $Y_{11,12}$. Above 400 Hz resonance peaks of the input mobility
$Y_{11,11}$ are flattened increasingly, as the bending wavelength
approaches twice the patch size (around $550\,{\rm Hz}$). 

In the high frequency range the patches are much larger than the structural
wave\-length and the averaged response of the plate due to uniformly
distributed pressure excitation, that is, the wave mobility of the
plate becomes mass governed \citep{Cremer_BOOK_2005}, a feature which
constitutes the basis for the well known mass-law for airborne sound
transmissions of panels. The relative magnitude of the transfer mobility
with respect to the input mobility decreases at higher frequencies. 

\begin{figure}
\begin{centering}
\includegraphics[width=0.8\textwidth]{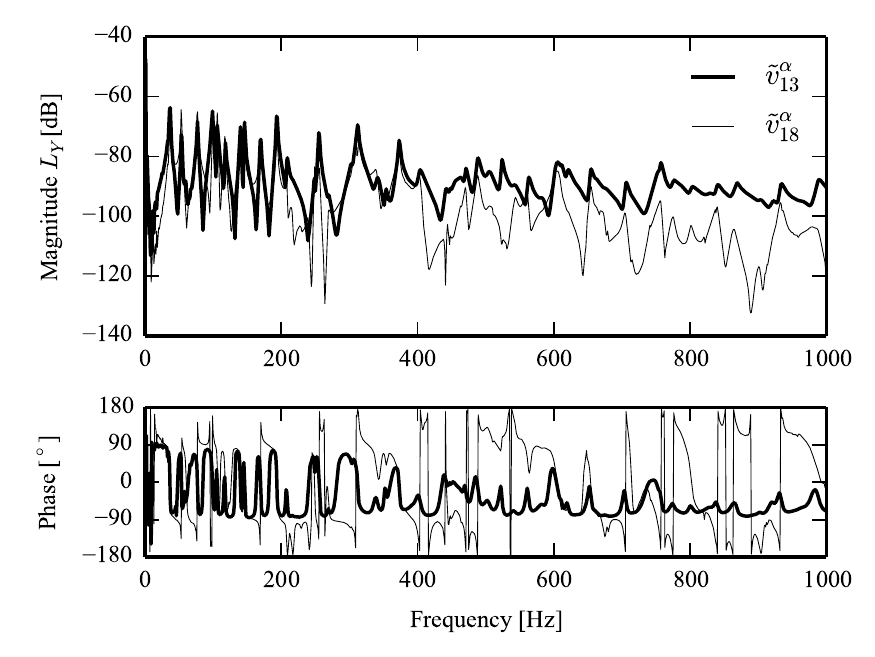}
\par\end{centering}

\protect\caption{Example of components of the bare plate velocity vector $\tilde{\mathbf{v}}^{\alpha}$
on patches 13 and 18. Here, $L_{Y}=20\log_{10}\left|\frac{v}{F}\right|$
is the response to a force excitation of $F=1\,{\rm N}$ at the centre
of patch 13.\label{fig:Yplate-1} }
\end{figure}

Figure \ref{fig:Yplate-1} shows an example of bare plate velocities
$\tilde{\mathbf{v}}^{\alpha}$ (source terms). The (internal) excitation
consisted of a point force of $1\,{\rm N}$ at the centre of patch
13. In the high frequency range the source terms exhibit a more pronounced
modal behaviour than the structural mobilities in Figure \ref{fig:Yplate}.
This is due to the different types of excitation -- uniform pressure
for the structural mobilities and point force for the source terms.

\subsection{Lining material characterisation\label{sub:Lining-material-characterisation}}

In this section some liner characterisation results are discussed.
In this case the chosen liner consists of a typical automotive car
floor trim package featuring three layers: a $7\,{\rm mm}$ moulded
polyurethane foam of density $85\,{\rm kg}/{\rm m^{3}}$ (structure
side) covered by a $2.2\,{\rm kg/m^{2}}$ polymeric heavy layer topped
by a $3\,{\rm mm}$ carpet (air cavity side).

\subsubsection{Bottom input/transfer impedance with open top, $\mathbf{h}^{\alpha\alpha}$}

\begin{figure}
\begin{centering}
\includegraphics[width=0.8\textwidth]{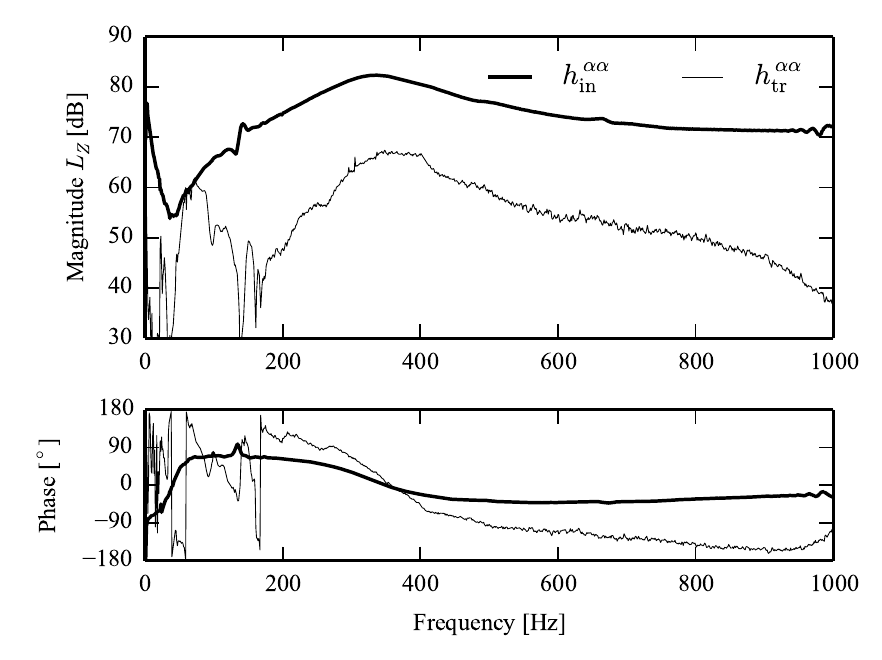}
\par\end{centering}

\protect\caption{Bottom side input and transfer impedance with open top, $\mathbf{h}^{\alpha\alpha}$.\label{fig:haa}}
\end{figure}

The sub-matrix $\mathbf{h}^{\alpha\alpha}$ of $\mathbf{H}$ contains
input terms $h_{{\rm in}}^{\alpha\alpha}=h_{11}^{\alpha\alpha}=h_{22}^{\alpha\alpha}$
and transfer terms $h_{{\rm tr}}^{\alpha\alpha}=h_{12}^{\alpha\alpha}=h_{21}^{\alpha\alpha}$.
As pointed out in section (\ref{sub:Patch-transfer-functions}) these
terms can be interpreted as the input impedance at the bottom surface
of the liner with pressure release conditions at its top surface.
The reconstructed $h_{{\rm in}}^{\alpha\alpha}$ and $h_{{\rm tr}}^{\alpha\alpha}$
curves are shown in Figure \ref{fig:haa}. 

Below $f=50\,{\rm Hz}$ the stiffness of the whole system, due to
excitation of a single patch and blocking the remaining patches, is
visible in the input term $h_{{\rm in}}^{\alpha\alpha}$. Between
$f=50\,{\rm Hz}$ and $f=300\,{\rm Hz}$ the input term shows mass-governed
behaviour, representing the mass of the full liner, that is, foam
plus heavy layer plus carpet. At $f\approx335\,{\rm Hz}$ the mass
of the heavy layer including the carpet resonates on the stiffness
of the foam layer. Above $400\,{\rm Hz}$ the heavy layer is decoupled
and the amplitude of $h_{{\rm in}}^{\alpha\alpha}$ shows a stiffness-like
decreasing trend with frequency. The contribution of $h_{{\rm tr}}^{\alpha\alpha}$
is most pronounced around the resonance. Below $f=100\,{\rm Hz}$
transfer term measurements are not reliable due to structural vibration
modes of the test rig and sensor noise.

\subsubsection{Cross/cross-transfer terms, $\mathbf{h}^{\alpha\beta}$ and $\mathbf{h}^{\beta\alpha}$}

\begin{figure}
\begin{centering}
\includegraphics[width=0.8\textwidth]{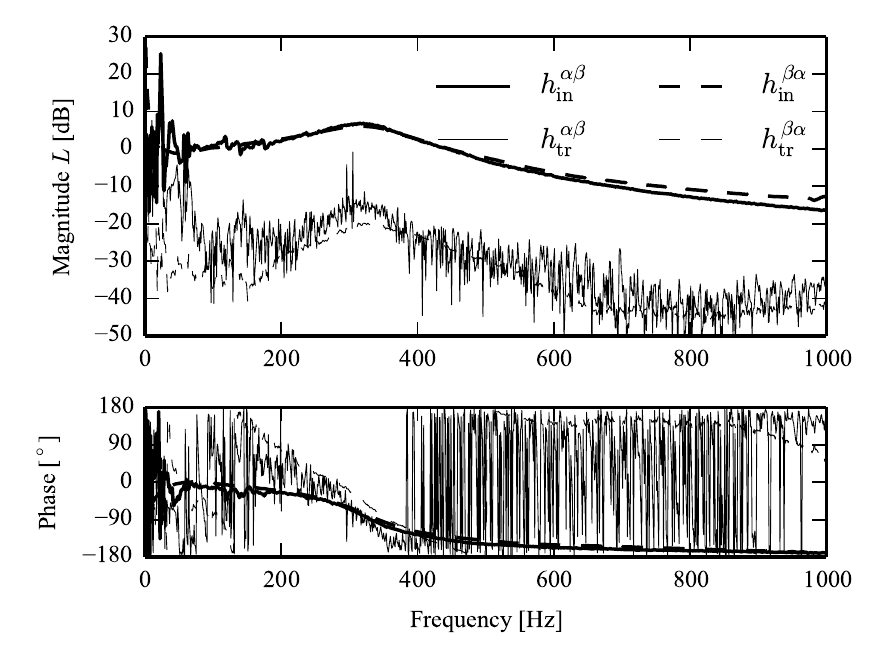}
\par\end{centering}

\protect\caption{Cross terms $\mathbf{h}^{\alpha\beta}$ (speaker excitation above)
and $\mathbf{h}^{\beta\alpha}$ (shaker excitation below).\label{fig:hab_hba}}
\end{figure}

Cross terms $\mathbf{h}^{\alpha\beta}$ and $\mathbf{h}^{\beta\alpha}$
are plotted in Figure \ref{fig:hab_hba}. Below $f=200\,{\rm Hz}$
the two cross-input terms $h_{{\rm in}}^{\alpha\beta}$ and $h_{{\rm in}}^{\beta\alpha}$
are close to unity. For the case of $h_{{\rm in}}^{\beta\alpha}$
this means that a velocity excitation on $\alpha$ would propagate
to $\beta$ with the same amplitude and phase if no pressure was exerted
on $\beta$. The same holds for $h_{{\rm in}}^{\alpha\beta}$ with
a propagation of pressure from $\beta$ to $\alpha$ with blocked
surface $\alpha$. This is expected, as the specimen should behave
like a rigid body below its eigenfrequencies. Errors due to structural
modes of the test rig are clearly visible below $f=100\,{\rm Hz}$.
Measurements of the pressure cross transfer term $h_{{\rm tr}}^{\alpha\beta}$
are limited by the signal-to-noise ratio. The terms show approximate
reciprocity, but are not identical for at least two reasons: Firstly,
reciprocity is only exactly valid for point transfer functions, but
not necessarily for patches. Secondly, in the given set-up there is
approximately a pressure-release condition on the top and a blocked
condition on the bottom side. On the bottom side skeleton and fluid
of the polyurethane foam are excited with the same velocity, whereas
on the top side the fluid in the carpet layer is excited more efficiently
by pressure than the stiffer and more massive skeleton (see section
\ref{sub:Bottom-surface-passive}).

\subsubsection{Top input/transfer mobility with blocked bottom, $\mathbf{h}^{\beta\beta}$}

\begin{figure}
\begin{centering}
\includegraphics[width=0.8\textwidth]{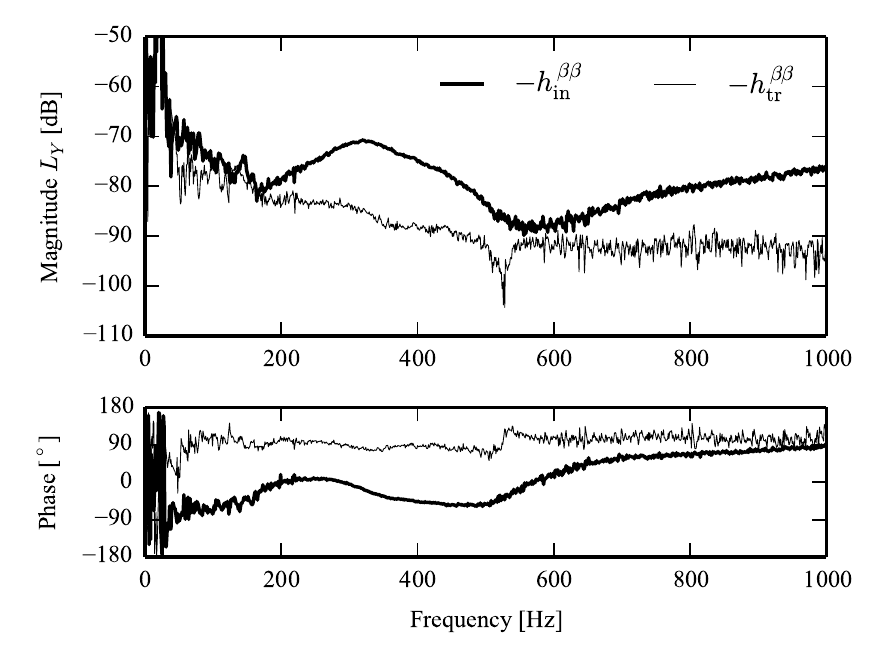}
\par\end{centering}

\protect\caption{Top mobility with blocked bottom, $\mathbf{h}^{\beta\beta}$ with
air gap correction applied.\label{fig:hbb}}
\end{figure}

Figure \ref{fig:hbb} shows the top mobility of the sample\footnote{Negative signs arise from the convention of positive velocities in
the upwards direction, see section \ref{sub:Patch-transfer-functions}.\label{fn:The-minus-sign}}. Errors related to the air gap during the characterisation have been
suppressed by the correction method described in section \ref{sub:Air-gap-correction}
.
The top mobility measurement is the only case where in addition to
the resonance also the antiresonance of the heavy layer-carpet system
is relevant. This antiresonance is visible at $f\approx550\,{\rm Hz}$
and the behaviour follows a spring-mass-spring like curve: Below the
resonance at $f\approx335\,{\rm Hz}$ the foam stiffness governs the
response. In the frequency range of $f=400-500\,{\rm Hz}$ between
resonance and antiresonance a mass-like behaviour of the decoupled
mass of heavy-layer plus carpet is visible. Above the antiresonance,
starting from $f={\rm 700}\,{\rm Hz}$, the mobility is mainly governed
by the carpet stiffness and damping. In this frequency range absorptive
effects of the carpet layer start to become relevant for the cavity
side. The accuracy of the results is limited by the air gap correction
in both low and high frequency range, where the measured signal is
determined mostly by the air gap impedance and not by the liner surface
impedance.

\subsubsection{Air gap correction}

\begin{figure}
\begin{centering}
\includegraphics[width=0.8\textwidth]{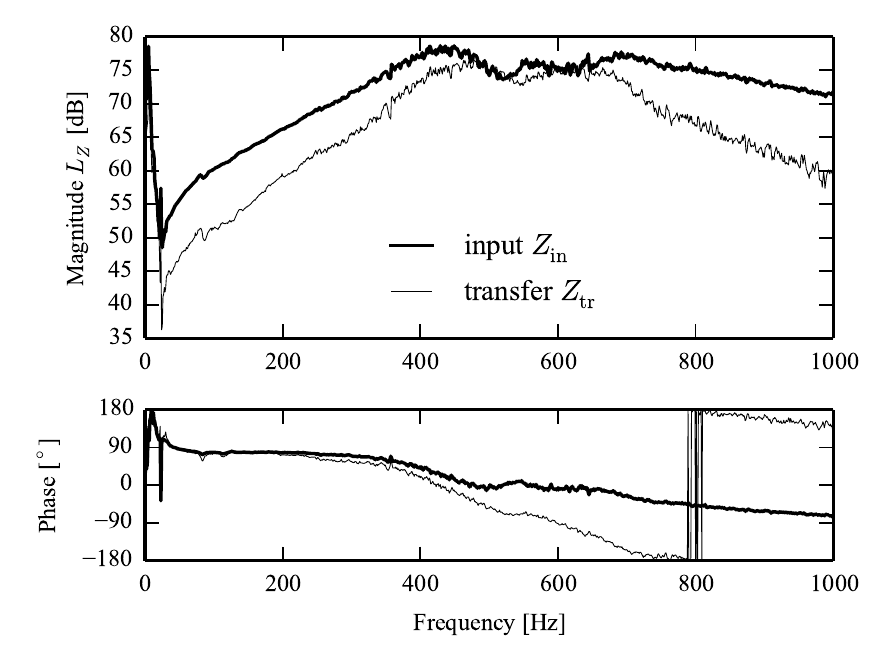}
\par\end{centering}

\protect\caption{Air gap input and transfer impedance (liner specimen replaced by a
rigid surface).\label{fig:airgap-impedance}}
\end{figure}

Figure \ref{fig:airgap-impedance} shows results of measured air gap
input and transfer impedances. \emph{PU}-probes were placed $5\,{\rm mm}$
from the specimen surface for this measurement.

\begin{figure}
\begin{centering}
\includegraphics[width=0.8\textwidth]{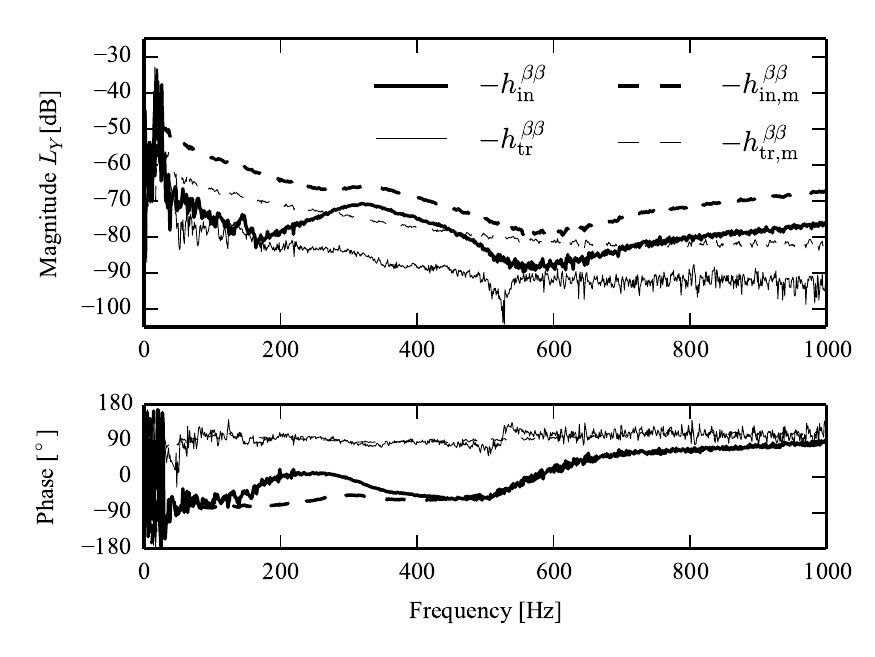}
\par\end{centering}

\protect\caption{Comparison of corrected ($h_{{\rm in}}^{\beta\beta}$ and $h_{{\rm tr}}^{\beta\beta}$)
and uncorrected ($h_{{\rm in,m}}^{\beta\beta}$ and $h_{{\rm tr,m}}^{\beta\beta}$)
surface input and transfer mobilities. \label{fig:airgap-impedance-1}}
\end{figure}

The air gap results indicate the behaviour described in \ref{sub:Air-gap-correction}:
In the frequency range below $400\,{\rm Hz}$ the result is governed
by the reactive part of the acoustic side wall radiation impedance.
With rising frequency, the finite stiffness of the air gap becomes
increasingly important. The correction eliminates the low-impedance
gap masking the specimen surface mobility (Figure \ref{fig:airgap-impedance-1}).

 The correction of $h_{{\rm in}}^{\beta\beta}$ is significant across
the whole frequency range. Especially in the stiffness dominated region
below the first liner resonance the mass-governed leakage through
the air gap edges is eliminated. The resonance of the heavy layer
on the foam, while barely visible in the uncorrected case, is represented
clearly in the corrected case. The transfer mobility $h_{{\rm tr}}^{\beta\beta}$
between the patches is reduced for all frequencies in the corrected
version and remains significant around the anti-resonance at $550\,{\rm Hz}$.
The correction is limited by measurement noise. Accurate placement
of the \emph{PU} array is critical for the correction due to the strong
particle velocity gradient near hard surfaces. 

It should be noted, that the top surface could also be characterised
with the interface defined on top of the air gap, that is, the air gap
could be included in the liner. Consequently, no correction would
be necessary. However, in this case further transfer terms would have
to be characterised in order to account for propagation inside the
air gap to more distant patches.

\section{Results for a coupled system}

An experimental case study for the described coupling method has been
performed on a physically coupled plate-liner-cavity system (Figure
\ref{fig:Coupled-system-discretised}) with patches of $20\times20\,{\rm cm}$.
The system consists of the $5\,{\rm mm}$ aluminium plate described
in section \ref{sub:Plate-characterisation}, a close-to-rigid cavity
of $1.7\times0.8\times1.0\,{\rm m}$ and a mass-spring type liner
($14\,{\rm mm}$ polyurethane foam, $\rho=69\,{\rm kg}/{\rm m}{}^{3}$
and a $3\,{\rm kg}/{\rm m}{}^{2}$ heavy layer). The liner was characterised
by the described procedure with similar results to the ones for the
liner with an additional carpet layer described in the previous section,
which was not available during the validation measurements. It was
nevertheless chosen to present characterisation results for the latter,
since it shows a richer behaviour in the frequency range where the
carpet acoustic absorption is relevant.

The analytical solution described in \citep{Pavic_CSV_2010} has been
applied to generate the cavity im\-pedance matrix. Results of the
investigations and an example for a detailed analysis of effects of
the liner on the system are presented below. For validation measurements,
sound pressure levels $L_{p}=20\,{\rm log}10\left|p/p_{0}\right|$
are normalised to $p_{0}=20\,\mu{\rm Pa}$.

\subsection{Expected interaction between structure, liner and cavity}

\begin{figure}
\begin{centering}
\includegraphics[width=0.8\textwidth]{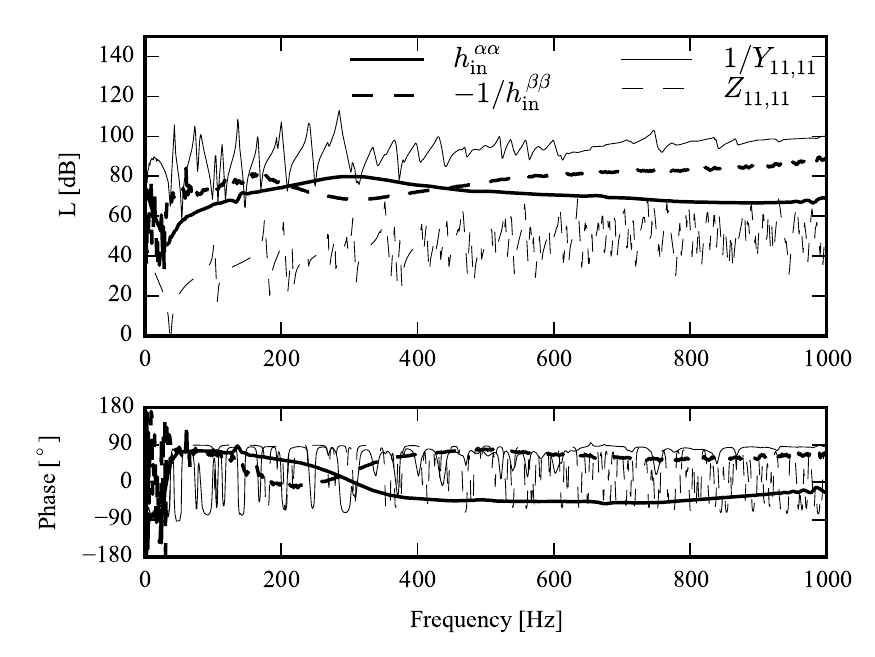}
\par\end{centering}

\protect\caption{Comparison of impedance of structure, liner and cavity. The solid
thin curve shows the inverse of a plate input mobility term, $1/Y_{11,11}$,
the dashed thin curve a cavity input impedance term, $Z_{11,11}$.
In-between, liner input impedances $h_{{\rm in}}^{\alpha\alpha}$
and $-1/h_{{\rm in}}^{\beta\beta}$are displayed. \label{fig:Comparison}}
\end{figure}

To illustrate regions where most interaction of the liner with the
adjacent structural and cavity domains occurs, representative examples
of interacting surface input terms are displayed in terms of impedances
in Figure \ref{fig:Comparison}.

From this picture most structure-liner interaction is expected at
the liner mass-spring resonance and at lower frequencies. Above the
sampling limit for the structure at $f=550\,{\rm Hz}$ plate mode
peaks are smeared out as mentioned in section \ref{sub:Plate-characterisation},
so the approach is not representative any more for coupling between
structure and liner. The sampling limit for air is given by $f=850\,{\rm Hz}$.
Strong interaction of the liner with the cavity is limited to the
range around the liner mass-spring resonance of $300\,{\rm Hz}$,
since no absorptive layer was included in this case study. Without
liner treatment plate and cavity could be considered as weakly coupled
in this frequency range, since their surface input impedances are
of different orders of magnitude.

\subsection{Validation and comparison to the bare system}

\begin{figure}
\begin{centering}
\includegraphics[width=0.8\textwidth]{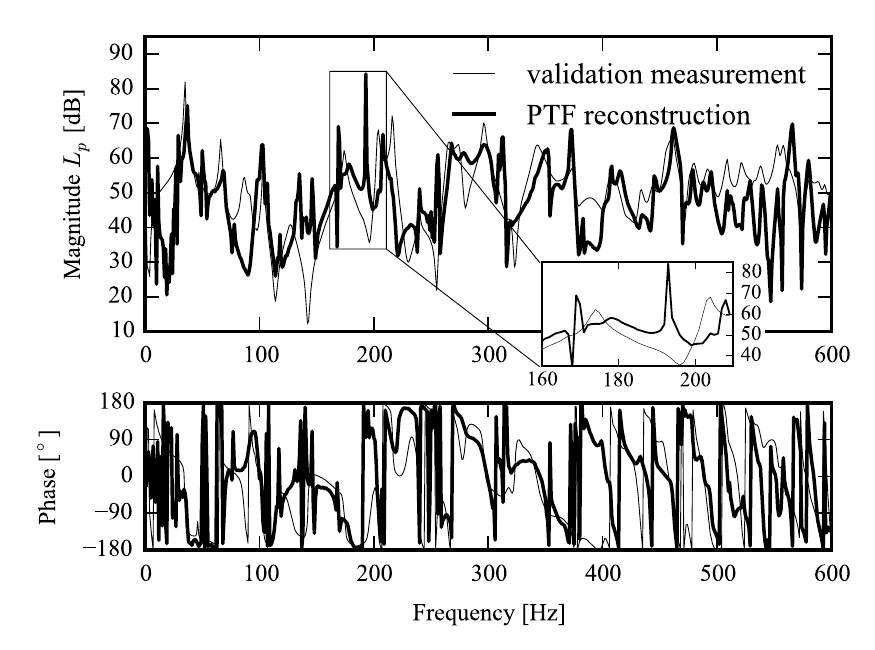}
\par\end{centering}

\protect\caption{Comparison of measured and reconstructed sound pressure level at the
reference microphone position in the cavity. A frequency region where
the reconstruction shows spikes is shown in a detailed sub-plot.\label{fig:validation}}
\end{figure}

To assess the coupling procedure's accuracy and validity range, the
sound pressure level at a reference position in the cavity due to
a force excitation of $F=1N$ in the centre of patch 13 (Fig. \ref{fig:Yplate-2})
is compared to the reconstructed value in Figure \ref{fig:validation}.

Except for isolated spikes at frequencies between $150\,{\rm Hz}$
and $200\,{\rm Hz}$ the trend of the measured signal is captured
by the reconstruction up to $500\,{\rm Hz}$. While certain modes
show a reasonable match in terms of amplitude and damping, differences
up to $20\,{\rm dB}$ in the amplitude and large phase differences
can be identified in several regions. The reason for the appearance
of the spikes is most likely related to inaccuracies in the characterisation
of the liner matrix elements $\mathbf{h}^{\alpha\alpha}$ (similar
to Figure \ref{fig:haa}) and/or the plate mobility matrix $\mathbf{Y}$.
However, this topic requires further investigations. Above $500\,{\rm Hz}$
strong discrepancies between measurement and reconstruction appear
which is in accordance with reaching the sampling limit for plate
bending waves at $550\,{\rm Hz}$.

\begin{figure}
\begin{centering}
\includegraphics[width=0.8\textwidth]{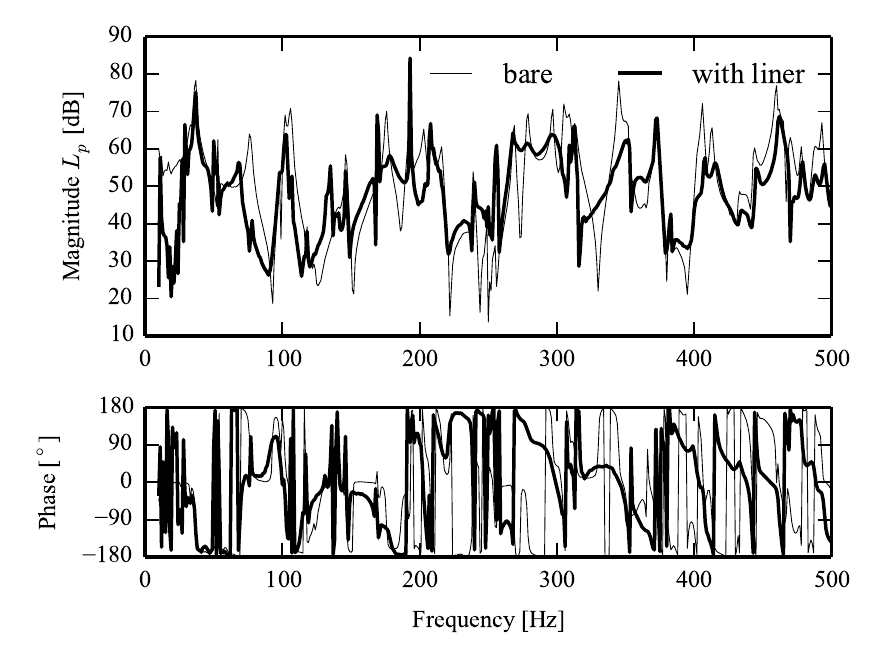}
\par\end{centering}

\protect\caption{Reconstructed sound pressure level at the reference position, comparison
to reconstruction of the bare system without liner.\label{fig:rec_bare}}
\end{figure}

Figure \ref{fig:rec_bare} shows a comparison of a reconstruction
of the bare plate radiating into the cavity and the system response
including the liner within the validity region.

Reduction in the sound pressure level is observed across the whole
validity range, reaching more than $10\,{\rm dB}$ at and above the
mass-spring resonance frequency. This is comparable to the reduction
observed in direct measurements. In the region below the resonance
plate modes are slightly shifted to lower frequencies due to additional
mass loading.

\subsection{Analysis of noise level reduction mechanisms}

\begin{figure}
\begin{centering}
\includegraphics[width=0.8\textwidth]{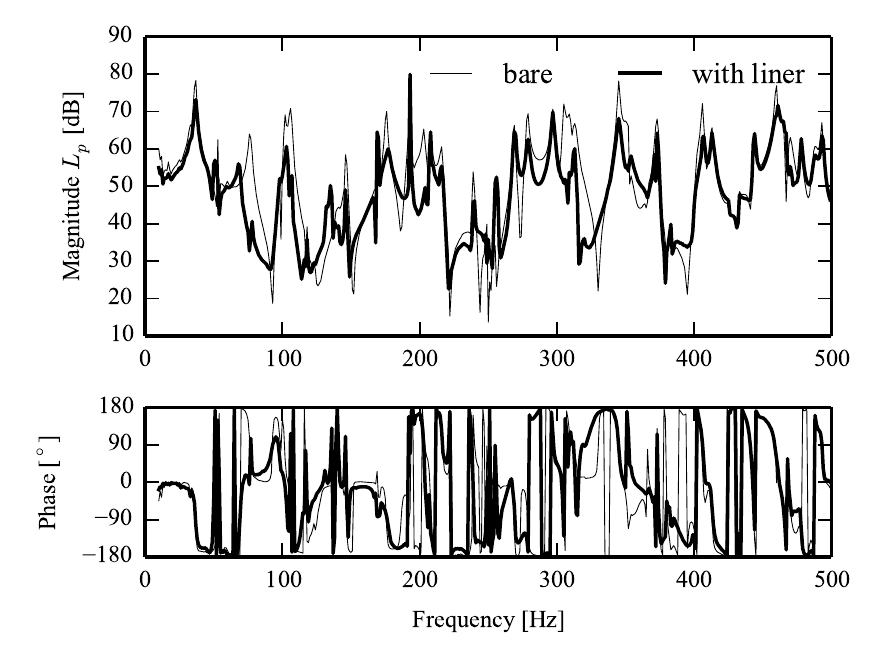}
\par\end{centering}

\protect\caption{Reconstructed sound pressure level at reference position -- considering
only plate loading $\mathbf{h}^{\alpha\alpha}$ ($\mathbf{h}^{\alpha\beta}=\mathbf{h}^{\beta\alpha}=\mathbf{I},\,\mathbf{h}^{\beta\beta}=\mathbf{0}$).\label{fig:Reconstructed-sound-pressure}}
\end{figure}

As mentioned in the introduction, inserting lining material reduces
the sound pressure level inside the cavity by a combination of structural
damping, transmission through the liner and cavity damping. The described
coupling me\-thod is well suited to analyse the respective contributions
from these different mechanisms. In order to study a certain aspect,
the corresponding matrix element is kept and the remaining ones are
fixed to artificially disable all other effects on the system. Figures
\ref{fig:Reconstructed-sound-pressure}-\ref{fig:Reconstructed-sound-pressure-1}
show the effect of the liner for the cases of only structural loading,
transmission or cavity loading respectively.

\begin{figure}
\begin{centering}
\includegraphics[width=0.8\textwidth]{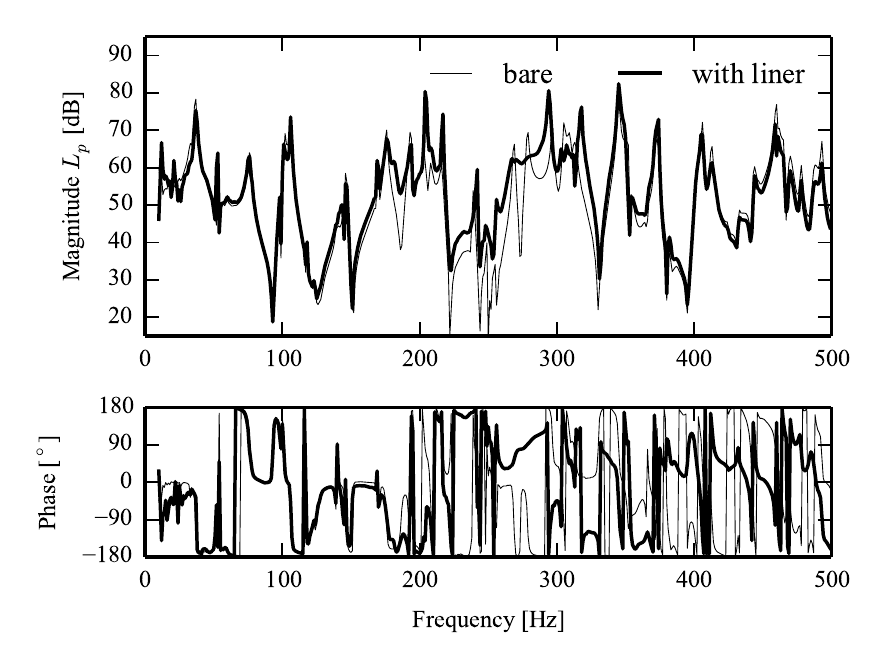}
\par\end{centering}

\protect\caption{Reconstructed sound pressure level at reference position -- considering
only transmission $\mathbf{h}^{\alpha\beta},\,\mathbf{h}^{\beta\alpha}$
($\mathbf{h}^{\alpha\alpha}=\mathbf{h}^{\beta\beta}=\mathbf{0}$).}
\end{figure}

\begin{figure}
\begin{centering}
\includegraphics[width=0.8\textwidth]{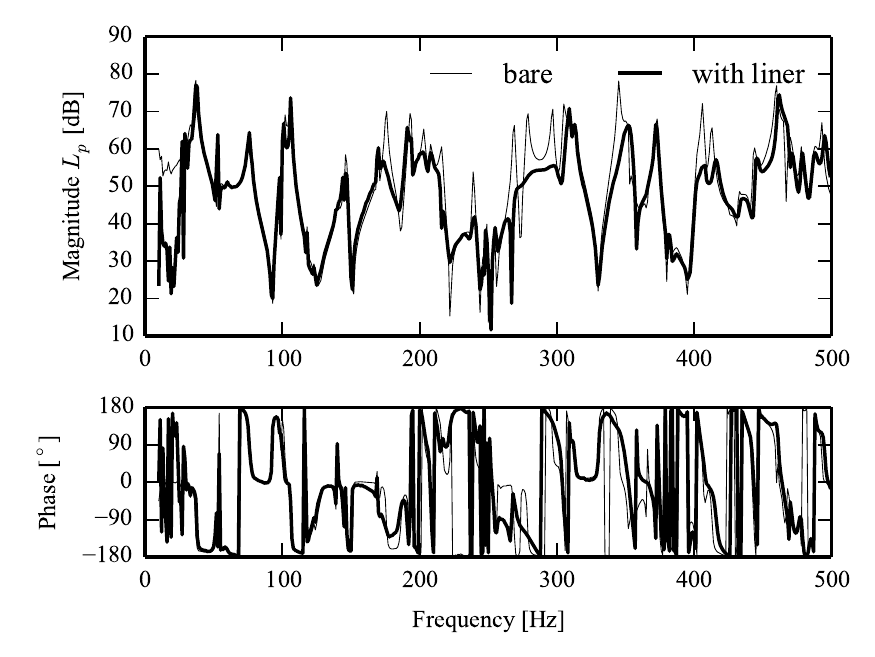}
\par\end{centering}

\protect\caption{Reconstructed sound pressure level at the reference position -- considering
only cavity loading $\mathbf{h}^{\beta\beta}$ ($\mathbf{h}^{\alpha\alpha}=\mathbf{0},\,\mathbf{h}^{\alpha\beta}=\mathbf{h}^{\beta\alpha}=\mathbf{I}$).
\label{fig:Reconstructed-sound-pressure-1}}
\end{figure}

Structural damping is dominant below and around the mass-spring resonance
of the liner at $300\,{\rm Hz}$ (Figure \ref{fig:Reconstructed-sound-pressure}).
At frequencies above $450\,{\rm Hz}$ little damping is observed.
Transmission through the liner is enhanced around the mass-spring
resonance frequency and is reduced by decoupling at higher frequencies.
In the overall result (Figure \ref{fig:rec_bare}) this leads to a
net increase of the cavity sound pressure level in some frequency
zones between $200\,{\rm Hz}$ and $400\,{\rm Hz}$. The remaining
spikes in the reconstruction are considered to be artefacts as mentioned
in the previous subsection. Cavity damping is realised mostly around
the mass-spring resonance. This seems reasonable, since the surface
of the heavy layer was airtight and should not act as a porous absorber.

\section{Limitations and Outlook\label{sec:Limitations}}

The coupling approach by patch transfer functions and the presented
characterisation methods are subject to several limitations that will
be summarised here. Most importantly, spatial discretisation leads
to an upper frequency limit caused by aliasing when the wavelength
approaches twice the patch dimension. For strongly interacting systems
the wavelength criterion is expected to be even stricter \citep{Aucejo_CS_2010}.
To extend the valid frequency range one needs to acquire transfer
functions between a larger number of smaller patches. Once the direct
experimental characterisation of the structure by the current method
becomes too time-consuming (section \ref{sub:Plate-characterisation}),
a numerical model would be preferred instead for this task. The same
holds for the proposed experimental characterisation of fluid domains.

In the coupling between structure and liner, shear stresses are not
considered explicitly (section \ref{sub:Patch-transfer-functions}).
While this approximation is justified for certain systems, the explicit
measurement of shear stresses may be required for liners that are
strongly bonded to a structure. 

As opposed to models based on material parameters, experimentally
characterised patch transfer functions do not allow for later adjustments
in material parameters or geometry. The assumption of rapid spatial
decay of transfer functions in the liner (section \ref{sub:Liner-characterisation---})
has to be verified for the respective specimen and patch dimension.
For this purpose a test rig with a movable patch configuration would
be advantageous. To obtain transfer functions to further patches one
might consider the enhancement of characterisation data by Green's
function methods \citep{Alimonti_JASA_2014}. A comparison to results
from material models based on Biot parameters would be of great interest
for cross-validation and to find possibilities to improve the liner
characterisation method.

\section{Concluding remarks}

A prediction method for the vibro-acous\-tic response of a coupled
system consisting of structure, lining material and fluid has been
presented. A patch transfer function approach employing experimentally
obtained subsystem characterisation data has been applied. Experimental
methods have been described for a plate and a lining material with
a heavy layer topped with a carpet. Results from liner measurements
indicate, that for typical multi-layer materials, a reasonably accurate
characterisation of input and next-neighbour terms at a length scale
of $20\,{\rm cm}$ is possible in a frequency range of $100-1000\,{\rm Hz}$.
The accuracy is mainly limited by resonances in the test rig, the
finite air gap and sensor signal-to-noise ratios. The predicted response
of a plate-liner-cavity-system was compared to validation measurements
performed on a physically coupled system. Input data for the reconstruction
consisted of experimentally determined impedance matrices for structure
and liner and an analytical model for an air cavity with hard walls.
Reconstructed results are comparable to results from validation measurements
in the expected validity region. As an application example, different
aspects of lining material effects on the coupled response (structural
damping, transmission, cavity damping) were analysed separately.

\section*{Acknowledgements}

The research work of Giorgio Veronesi has been funded by the European
Commission within the ITN Marie Curie Action project GRESIMO under
the 7th Framework Programme (EC grant agreement no. 290050). The authors
acknowledge the financial support of the ``COMET K2 \textendash{}
Competence Centres for Excellent Technologies Program\-me'' of the
Austrian Federal Ministry for Transport, Innovation and Technology
(BMVIT), the Austrian Federal Ministry of Science, Research and Economy
(BMWFW), the Austrian Research Promotion Agency (FFG), the Pro\-vince
of Styria and the Styrian Business Promotion Agency (SFG). Furthermore,
the authors express their gratitude to the consortium partners BMW
AG, IAC GmbH, Microflown Technologies, ESI GmbH and Université de
Sherbrooke for their support. Finally, the authors gratefully acknowledge
the support of COST action TU1105. 

\bibliographystyle{elsarticle-num-names}
\bibliography{characterisation}

\end{document}